\documentstyle[epsf,prl,twocolumn,aps]{revtex}
\begin{document}
%\draft
%\twocolumn[
\title{Universal cross-over behavior of a magnetic impurity 
and consequences for doping in spin-1/2 chains}
\author{Sebastian Eggert and Stefan Rommer}
\address{Institute of Theoretical Physics,
Chalmers University of Technology and G\"oteborg University,
S-412 96 G\"oteborg, Sweden}
\maketitle
\date{\today}
\begin{abstract}
%\widetext\leftskip=0.10753\textwidth \rightskip\leftskip
We consider a magnetic impurity in the antiferromagnetic spin-1/2 
chain which is equivalent to the two-channel Kondo problem in terms 
of the field theoretical description.  Using a modification of
the transfer-matrix density matrix renormalization group (DMRG)
we are able to determine local and global properties in the
thermodynamic limit.  The cross-over function for the 
impurity susceptibility is calculated over a large temperature range,
which exhibits universal data-collapse. We are also able to 
determine the local susceptibilities near the impurity, which 
show an interesting competition of boundary effects.  This 
results in quantitative predictions for experiments on doped
spin-1/2 chains, which could observe two-channel Kondo physics
directly.
\end{abstract}
%] \narrowtext
%\newpage
Magnetic impurities in low-dimensional antiferromagnets are recently 
of much theoretical %\cite{martins}
and experimental interest in connection with high
temperature superconductivity.  We now study an impurity model in the
spin-1/2 chain consisting of 
two altered bonds in the chain, which is known to have
an equivalent field theory description to the spin-sector of the 
two-channel Kondo (2CK) model\cite{eggert1,clark}.  
The 2CK model has received much interest in the
theoretical physics community\cite{review} since it was
first proposed in 1980\cite{blandin} and this model is often cited as 
a standard example of non-Fermi-liquid physics.  
Bethe ansatz\cite{review} and conformal field theory techniques\cite{affleck}
have lead to almost complete understanding. On the other hand it is
much less clear to what extend experimental applications exist,
although there is some hope that the conductance
behavior through metal constrictions can be explained by the 
2CK effect\cite{ralph}. 

To study the 2CK effect experimentally, site-parity symmetric bond 
defects could feasibly be created by doping quasi one-dimensional
spin-1/2 compounds, and Knight shift experiments would then be able to
observe 2CK physics explicitly.  Our goal is therefore
to provide quantitative predictions for the local susceptibilities
in a range around the impurity, which turn out to exhibit an interesting 
competition between two boundary effects.  Moreover, we are able
to explicitly show the expected data collapse of the impurity susceptibility
for different coupling strengths, and calculate the resulting 
cross-over function
over a large temperature range.  For this purpose we have developed a 
modification of the transfer-matrix DMRG\cite{TMDMRG}. 

The model we are considering is the antiferromagnetic spin-1/2 chain
with two altered bonds
\begin{equation}
H = J \sum_{i=1}^{L-1} \vec S_i \cdot \vec S_{i+1} \  +  \
J'(\vec S_L  + \vec S_1 ) \cdot \vec S_0 , \label{ham}
\end{equation}
which is known to have an equivalent field theory description of the 
2CK effect as we will outline below.  Interestingly, an
integrable spin-1 chain with a spin-1/2 impurity is also known to be 
equivalent to the 2CK problem on the level of the Bethe ansatz 
equations\cite{schlottmann}, but we are not able to analyze that model in terms
of the field theory.

The low-temperature and long-wavelength properties of the
unperturbed spin-1/2 chain
are well described by a conformal field theory Hamiltonian in terms 
of 1+1 dimensional bosons  
\begin{equation}
 H = \int dx \ \frac{v}{2} \left[ (\partial_x \phi)^2 + \Pi_\phi^2 \right],
\label{bosham}
\end{equation}
where $\Pi_\phi$ is the conjugate operator to the boson field $\phi$.
This field theory has been discussed in more detail elsewhere\cite{eggert1}, 
and we will just focus on a more pedagogical description in this Letter.
For temperatures and wave-vectors well below some cutoff $\Lambda \sim J$ 
the spin-1/2 chain exhibits critical behavior and
scale invariance. Scale-invariance means that we get the same physical results
for some quantity $\cal O$
if we rescale the temperature or distances with some factor $\Gamma$,
as long as we also multiply the physical quantity $\cal O$ with $\Gamma^d$,
\begin{equation}
{\cal O}(T) = \Gamma^{-d} {\cal O}(\Gamma T),
\end{equation}
where $d$ is referred to as the scaling dimension of $\cal O$, and the
temperature $T$ can also be replaced by the inverse system size $1/L$.
For example, operators in the ``free'' Hamiltonian (\ref{bosham}) 
have a scaling dimension of $d=1$ (after integration over $x$). 
This means that the energy spacing of the spectrum 
is proportional to $1/L$.  ``Higher order operators'' in the Hamiltonian
are operators with $d>1$, which give small corrections to the spectrum 
of order $1/L^d$ and are hence termed ``irrelevant'' and are neglected
in Eq.~(\ref{bosham}).
(We have also neglected one marginal operator $\cos \sqrt{8 \pi}\phi$
which turns out to be justified although the corrections are only 
logarithmically small). For the rest of this Letter we will work in the 
thermodynamic limit $L\to \infty$ and rescale $T$ instead.

We can now analyze the perturbation $J'$ on the bonds in Eq.~(\ref{ham}) 
in terms of the scaling dimensions of the local operators, which
arise in Eq.~(\ref{bosham}) due to the broken translational invariance.
If the perturbation on the bonds 
$\delta J = J-J'$ is small, we can use perturbation theory on
a chain with periodic boundary conditions.  In this case the local operator
with the lowest scaling dimension which still observes the site-parity
symmetry of the problem is known to be $\partial_x \sin{\sqrt{2 \pi} \phi}$
of dimension $d=3/2$\cite{eggert1}.  This
operator gives the leading corrections, but is still irrelevant.
In particular, for a given coupling strength $\delta J$ the size of the 
corrections becomes effectively smaller by a factor 
of $\Gamma^{d-1}$ if we rescale $T$ by $\Gamma < 1$.
The opposite is also true:  If we increase the temperature the 
effective perturbation strength $\Gamma^{d-1} \delta J$
may become so strong that a systematic expansion
fails at some special temperature $T_K$, called the cross-over (or Kondo) 
temperature.  Above $T_K$ we therefore expect a completely
different behavior, namely that of an open chain if $T_K < T < \Lambda$.  
We say that the system 
{\it renormalizes} from an open boundary condition 
to the infrared fixed point of a healed periodic chain as the 
temperature is lowered. For small $\delta J$ we have {\it defined}
$T_K$ as the temperature at which the product $T_K^{d-1} \delta J$ 
becomes large, so that we can write
$T_K \propto \delta J^{1/(1-d)} = \delta J^{-2}$ in this limit. 
Hence, rescaling $T$ by $\Gamma$ is equivalent to changing $T_K$ 
by $1/\Gamma$, i.e.~altering  the initial coupling strength $\delta J$
by $\Gamma^{d-1}= \Gamma^{1/2}$, which is really the meaning of 
renormalization.

To compare this system with the 2CK effect it is more
instructive to start with open boundary conditions and 
consider a weak antiferromagnetic coupling $0< J' \ll J$.
In this case
the leading operator is $S^z_0 [\partial_x \phi(0) + \partial_x \phi(L)]$
with dimension $d=1$ which turns out to give logarithmically {\it relevant}
contributions as the temperature is lowered.   For small $J'$, 
the Kondo-temperature
is given by $T_K \propto e^{-b/J'},$ where $b$ is some constant.
If we identify the central spin $\vec S_0$ with the Kondo-impurity and the
two ends of the chain with the spin-sectors of the two electron
channels, it is evident how this scenario is equivalent to the
2CK problem\cite{affleck}: 
A small coupling ($J' \ll J$) to the impurity is
marginally relevant as we lower the temperature 
and the system renormalizes to an intermediate
coupling ($J' = J$) fixed point, i.e.~the periodic chain.

We have chosen the numerical transfer matrix DMRG\cite{TMDMRG} to test
these concepts.  The partition function of the 
spin-1/2 chain can be written in terms of transfer matrices 
$Z = {\rm tr}({\cal T}^{L/2}) \ 
\stackrel{L \to \infty}{\longrightarrow}  \ \lambda^{L/2},$
where $\lambda$ is the largest eigenvalue of $\cal T$.  
The transfer matrix DMRG
constructs a transfer matrix for a finite number of time-slices $M$ and
then successively increases $M$ by keeping only the most relevant 
basis states that are necessary to calculate $\lambda$.  
We now extend this method to non-uniform systems
by using the eigenstate $|\psi_\lambda\rangle$ 
corresponding to the highest 
eigenvalue $\lambda$. We can include any impurity interaction 
which is described by a local matrix $\cal T_{\rm local}$
explicitly in the partition function 
\begin{equation}
Z = {\rm tr} 
({\cal T}^{L/2-1}{\cal T}_{\rm local}) \ 
\stackrel{L \to \infty}{\longrightarrow}  \ 
\lambda^{L/2-1} \langle \psi_\lambda|{\cal T}_{\rm local} 
|\psi_\lambda \rangle.
\end{equation} 
If the eigenstate $|\psi_\lambda\rangle$
is known to reliable precision from the DMRG, 
it is straightforward to determine any thermodynamic 
property from this expression (even locally).  
This method proved to be superior 
in speed, accuracy, and temperature range compared to Monte Carlo
simulations, which we have used for checking purposes.

\begin{figure}
\begin{center}
\mbox{\epsfxsize=3.35in \epsfbox{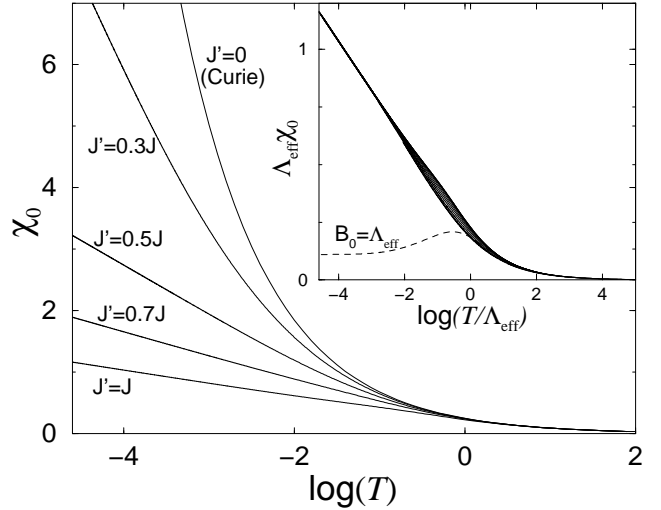}}
\end{center}
\caption{The linear response of the impurity spin to a local magnetic field.
Inset: the data-collapse for 14 different $J' = 0.1J,\ ...,\ J$ and
the effect of a {\it finite} field $B_0=\Lambda_{\rm eff}$.
\label{chilocal}}
\end{figure}
The first task is to show that our renormalization picture is 
accurate i.e.~that we can indeed use periodic boundary 
conditions to describe the system at sufficiently low temperatures.
For this purpose we consider  
the linear response $\chi_0$ of the impurity spin $\vec S_0 $ to a {\it local}
magnetic field $B_0$, which is given by the Kubo formula
\begin{equation}
\chi_0 (T) = \int^{1/T} \langle S^z_0(\tau) S^z_0(0)\rangle \, d \tau.
\label{kubo}
\end{equation}  
For open boundary conditions the impurity spin will be 
asymptotically free and the response is given by 
a Curie law  $\chi_0 \propto 1/ 4 T$. On the other hand, for periodic
boundary conditions the auto-correlation functions will be determined
by  the leading field theory operator which obeys
the same symmetry transformations as $\vec S_0$.
We found this operator to be
$\cos \sqrt{2 \pi} \phi$ with dimension $d=1/2$ and an
auto-correlation function proportional to $1/\tau$,
which should lead to a logarithmic divergence as $T\to 0$.
As can be seen in Fig.~(\ref{chilocal}), we indeed find
a cross-over from Curie law to logarithmic behavior at
an effective cutoff $\Lambda_{\rm eff} = {\rm min}(\Lambda, T_K)$,
depending on the coupling strength $J'$ with a scaling behavior of
the form $\chi_0(T) = 
g(T/\Lambda_{\rm eff})/\Lambda_{\rm eff}$ (inset).
The logarithmic behavior was observed 
before at one coupling only by other methods\cite{zhang}.  Note, 
however, that the response $\chi_0$
is only an indication of the auto-correlation functions, but it must not
be confused with the impurity susceptibility, since we have only applied 
the magnetic field at one spin. In fact every spin in the periodic chain 
shows a logarithmic response to a {\it local} magnetic field.

\begin{figure}
\begin{center}
\mbox{\epsfxsize=3.35in \epsfbox{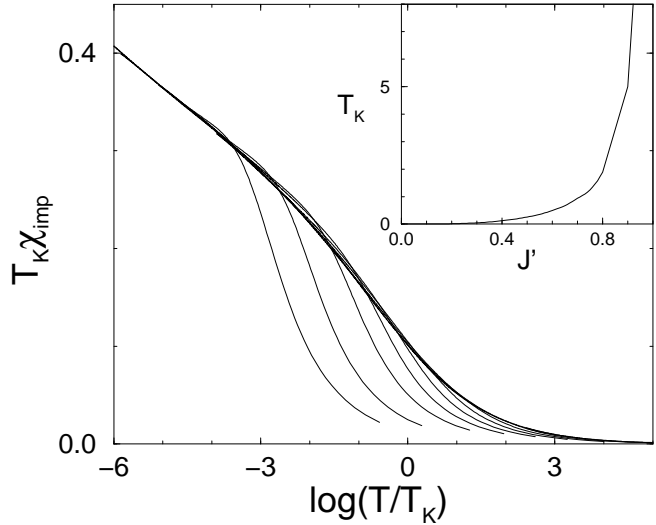}}
\end{center}
\caption{The scaled impurity susceptibility $T_K \chi_{\rm imp}$
for an appropriate 
choice of $T_K$ as a function of $J'= 0.1J, \ ...,\ 0.95J$ (inset).
\label{crossover}}
\end{figure}
From an experimental point of view it is much more interesting to look at
the impurity susceptibility, which can be defined as the size independent
contribution to the total system susceptibility 
$\chi_{\rm imp} = \lim_{L\to \infty} (\chi_{\rm system} - L \chi),$
where $\chi$ is the susceptibility per site far away from any boundary.
While our method is in principle capable of extracting the impurity 
susceptibility directly,
we were able to obtain results at much lower temperatures 
by considering the linear 
response of the impurity spin to a {\it uniform} field $B$ instead
\begin{equation}
\chi_{\rm imp} \approx \frac{d \langle S^z_0 \rangle_B}{d B}  - J' \chi,
\label{imp}
\end{equation}
which gives a good indication of the true impurity susceptibility 
(in contrast to $\chi_0$ in
Eq.~(\ref{kubo}), where we only considered a {\it local} field $B_0$).
From the 2CK effect it is known that the impurity susceptibility
is logarithmically divergent below $T_K$ while it shows a Curie-law behavior
above\cite{review,affleck}. Interestingly, this cross-over shows a 
universal data collapse, because changing the coupling strength
(i.e.~$T_K$) is equivalent to rescaling the temperature, i.e.~there
is only one independent variable $T/T_K$
\begin{equation}
\chi_{\rm imp}(T) = f(T/T_K)/T_K, \label{scaling} 
\end{equation}
where $f(x)$ is a universal function (ignoring higher order operators).
This data collapse is clearly seen in Fig.~(\ref{crossover}) with an 
appropriate choice
of $T_K$  as a function of $J'$ (inset), showing the predicted logarithmic 
scaling at low $T$.  The non-universal 
deviations of some of the curves at higher $T/T_K$  
are due to the fact that in those cases $T_K$ was so large that 
regions above the cutoff have been included.

In principle, a similar logarithmic scaling should be observable
for the impurity specific heat $C_{\rm imp}/T$.
However, we find that the critical scaling of the specific heat occurs 
at lower $T$ and instead $C_{\rm imp}$ shows a more complex
behavior in the intermediate range.  Moreover, the numerical method is known
to produce larger errors for $C$ at low $T$\cite{TMDMRG}, which may be due
to the second derivative involved.  Therefore, we could not fully reproduce 
the corresponding data collapse of $C_{\rm imp}/T$, but our data is nonetheless 
consistent with a cross-over to logarithmic scaling as $T\to 0$.
Just like in the 2CK effect we also find that a
{\it finite} magnetic field  (local or global) will change the logarithmic 
non-Fermi-liquid behavior and produce a cross-over to a constant susceptibility
below some temperature $T_{\rm FL}$ as shown in inset of Fig.~(\ref{chilocal}).
The different renormalization behavior can be traced to the broken spin-flip
symmetry, which allows the relevant operator $\cos \sqrt{2 \pi} {\phi}$
with $d=1/2$ at the periodic chain fixed point. This will result in a
quadratic field dependence $T_{\rm FL}\propto B_0^2$, which has 
already been demonstrated in Ref.~\onlinecite{review}.
In experiments, however, fields are expected to be small compared to $J$
and $T_K$, 
so that in most cases only the non-Fermi liquid behavior will be observed.
The effect of a finite $B_0$ is analogous to 
violating the symmetry condition in the resonant tunneling scenario for 
electrons\cite{kane}.  

Finally, we would like to consider the local susceptibilities (the Knight 
shift) of the individual spin sites in a region around the impurity as a 
function of site index $x$
\begin{equation}
\chi_{\rm local}(x) \ = \ \left.\frac{d \langle S^z(x)\rangle_B}{dB}
\right|_{B=0}
\end{equation}
In Ref.~\onlinecite{eggert2} the most dramatic effect of an open 
boundary condition on the Knight shifts was
found to be a {\it staggered} component 
$\chi_{\rm open}=\chi_{\rm local}-\chi$
in the response to a {\it uniform}
magnetic field $B$, which actually increases with site index $x$ 
\begin{equation}
\chi_{\rm open}(x) 
\ \propto \ (-1)^{x} \frac{x \sqrt{T}}{\sqrt{\sinh 4 x T}},
\label{chiopen}
\end{equation}
where $T$ is measured in units of $J$.

The staggered part in Eq.~(\ref{chiopen}) arises due to open boundary 
conditions and hence it will be diminished 
as the system renormalizes to the periodic chain fixed point.
However, there will be a whole new effect due to the leading 
irrelevant operator $\partial_x \sin{\sqrt{2 \pi} \phi}$ at the 
periodic chain fixed point. This operator also
induces a staggered part $\chi_{\rm periodic}$, 
but with opposite sign, i.e.~the induced
response at the first site $\vec S_1$ is {\it negative}.
The alternating response as a function of site index $x$ is now 
\begin{eqnarray} 
\chi_{\rm periodic} (x)& = & {(-1)^{x}}\frac{1}{T}\int\!\! dy\
\langle S^z_{\rm alt}(x)   S^z_{\rm uni}(y) \rangle \nonumber \\
 & \propto & (-1)^{x}\frac{1}{T}
 \int\!\! dy \int^{1/T} \!\! d\tau\, g(x,y,\tau),
\end{eqnarray} 
where $S^z_{\rm alt}$ and $S^z_{\rm uni}$ refer to the 
leading operators which describe the alternating ($\cos \sqrt{2 \pi} \phi$)
and uniform ($\partial_x \phi$) parts of the spin z-component, 
respectively.  The correlation function $g(x,y,\tau)$ is therefore given by
\begin{eqnarray} 
 g(x,y,\tau) & \propto & \langle 
\cos{\!\sqrt{2 \pi} \phi}(x) \, \partial_x \phi(y) \,
\partial_x \sin{\!\sqrt{2 \pi} \phi}(0,\tau) \rangle \nonumber \\
& \propto & \frac{T}{\sin{ T (\pi \tau - i 2 x)}} \frac{T^2}{\sin^2{T (\pi
\tau - i2 y)}},
\end{eqnarray}
where we have used standard field theory techniques (e.g.~the boson mode 
expansion\cite{eggert4}).
The integral of the second factor over the spatial coordinate $y$ is the 
same that determines the unperturbed susceptibility per site\cite{eggert3} 
and simply gives a $\tau$-independent contribution proportional to $T$.  
The integral of the first factor 
over the imaginary time variable can then easily be done, giving
\begin{equation} 
\chi_{\rm periodic}(x) \ \propto \ (-1)^x \log{\left[
\tanh{ (x T)}\right]}.
\label{chiper}
\end{equation}
This expression shows the logarithmic divergence with $T$
explicitly for small $x$ at the impurity, and the response then drops
off with $\exp{(-2 x T)}$ as $x T \to \infty$ (i.e.~with the
same exponential as observed for the open chain response (\ref{chiopen})). 

\begin{figure}
\begin{center}
\mbox{\epsfxsize=3.35in \epsfbox{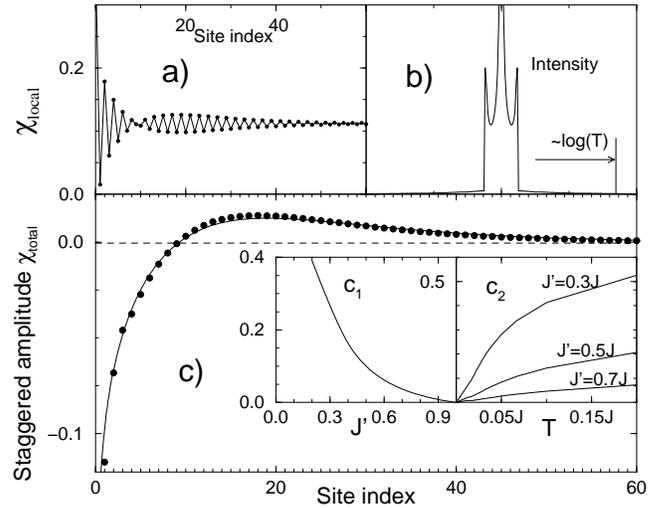}}
\end{center}
\caption{a.) The local susceptibilities as a function of site index 
for $T=0.05J$ and $J' = 0.7J$.
b) The corresponding
typical NMR spectrum with a distinct feature (kink).
c) The fit of the alternating amplitude to Eq.~(\ref{chi}) (solid line)
with the appropriate coefficients (inset). 
\label{knightshift}}
\end{figure}

As $J'$ is increased, the alternating part changes
from the behavior of Eq.~(\ref{chiopen}) to the behavior of the
stable fixed point in Eq.~(\ref{chiper}), which
is always logarithmically dominant as $T \to 0$. 
However, even very close to the periodic chain fixed point
we observe an interesting and complex 
competition of {\it both} contributions 
(see Fig.~(\ref{knightshift}a)), which is nonetheless 
completely understood.  In particular, below $T_K$
 the total amplitude of the
staggered part of $\chi_{\rm local}$ always fits very well to a superposition
\begin{equation}
\chi_{\rm total}(x) \ = \ c_1 \log[\tanh(x T)] 
\  + \ c_2 \frac{x \sqrt{T}}{\sqrt{\sinh 4 x T}}. \label{chi}
\end{equation}
This formula gave excellent results as can be seen for a typical fit 
in Fig.~(\ref{knightshift}c), and the coefficients have been 
determined for all values of $J' \ge 0.2J$ and $T$ (inset).  The coefficient 
$c_1$ is $T$-independent, while $c_2$ renormalizes to zero as $T\to 0$.

To see these effects experimentally, it will be 
necessary to induce site-symmetric bond defects into quasi one-dimensional 
spin-1/2 compounds like $\rm Sr_2CuO_3$, $\rm KCuF_3$, or Copper Pyrazine
Nitrate.  A simple doping with
another effective spin-1/2 ion for $\rm Cu^{2+}$ may be possible, 
but more likely the surrounding 
non-magnetic ions can be doped in order to create a suitable lattice 
deformation. The effect of open boundaries
has already been seen in NMR experiments\cite{NMR}, by observing a unique 
feature that broadened with $1/\sqrt{T}$ and had the predicted shape derived
from Eq.~(\ref{chiopen}).  For site-symmetric perturbations we
can now predict a broadening with $\log T$ of the 
NMR spectrum.  In addition there will be a feature (kink) 
at smaller Knight shifts which comes from the relative 
maximum in the alternating response (see Fig.~(\ref{knightshift}b)).  
This feature has a distinctive shape and temperature dependence
which can be calculated from the constants $c_1$ and $c_2$ with the
help of Eq.~(\ref{chi}) for any coupling strength $J'$. 
Ordinary susceptibility  measurements should be able to identify the
impurity contribution as predicted by the cross-over function in 
Fig.~(\ref{crossover}).
In a more exotic twist we can even imagine $\mu$SR experiments, where the
muon itself may play the role of an impurity, i.e.~depending on the
preferred muon location in the lattice, the muon could feasibly
induce a site-symmetric lattice distortion and at the same time
measure the logarithmically divergent response at the impurity site.

In summary, we have shown the cross-over of a two-channel Kondo
impurity explicitly,  which confirmed the renormalization picture
and demonstrated the expected data collapse explicitly.  Moreover,
we were able to predict 
the response to a uniform magnetic field of individual spin sites 
in a large region around the impurity, which led to quantitative 
predictions for Knight shift experiments on doped spin-1/2 compounds.

\begin{acknowledgements}
We would like to thank I.~Affleck, M.~Crawford, D.~Gustafsson, H.~Johannesson, 
D.~Johnston, A.~Kl\"umper, C.P.~Landee,
S.~\"Ostlund, I.~Peschel,  N.~Shibata, M.~Takigawa, and X.~Wang
for valuable contributions. 
This research was supported in part by the Swedish Natural Science 
Research Council (NFR).
\end{acknowledgements}

\end{document}